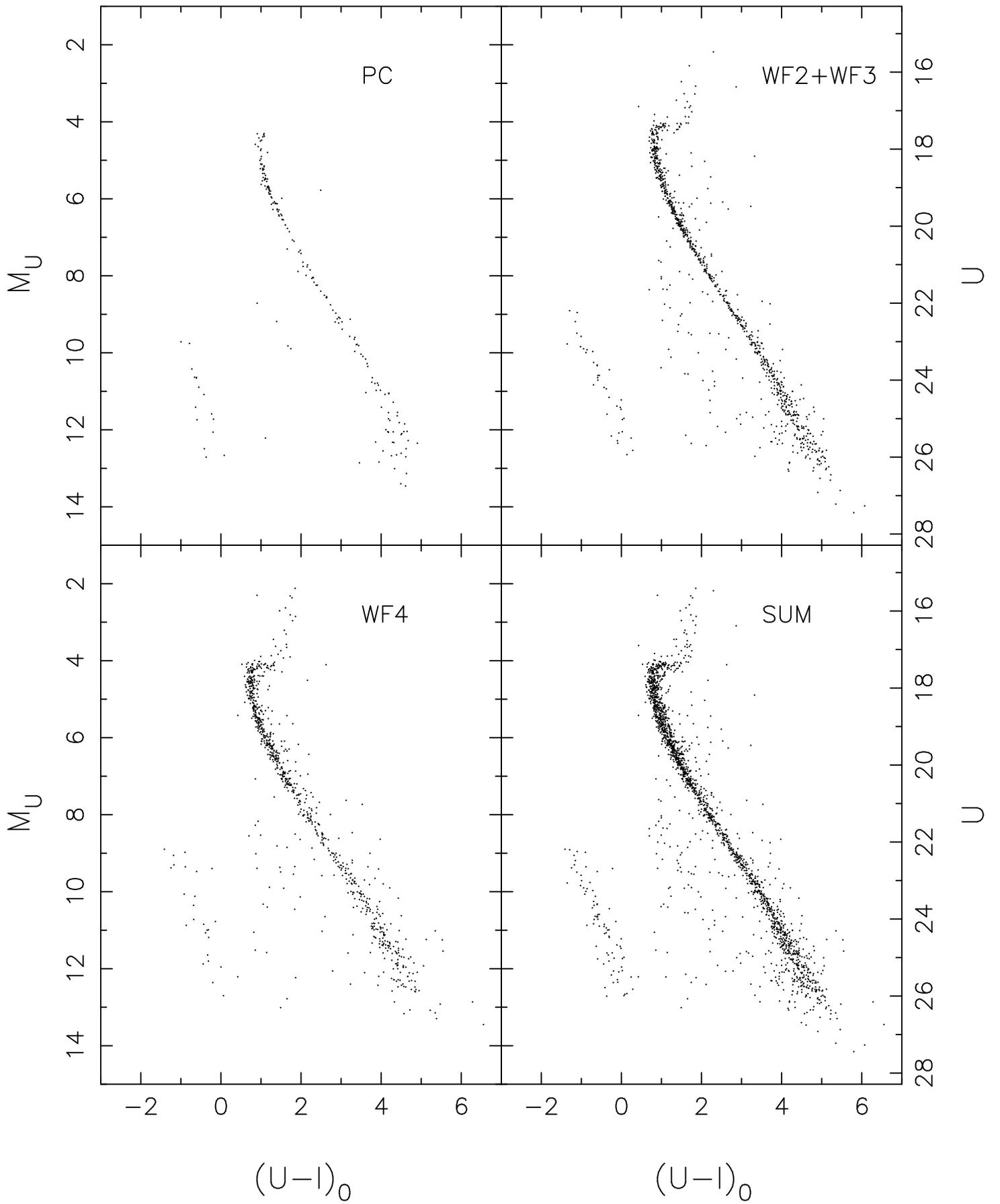

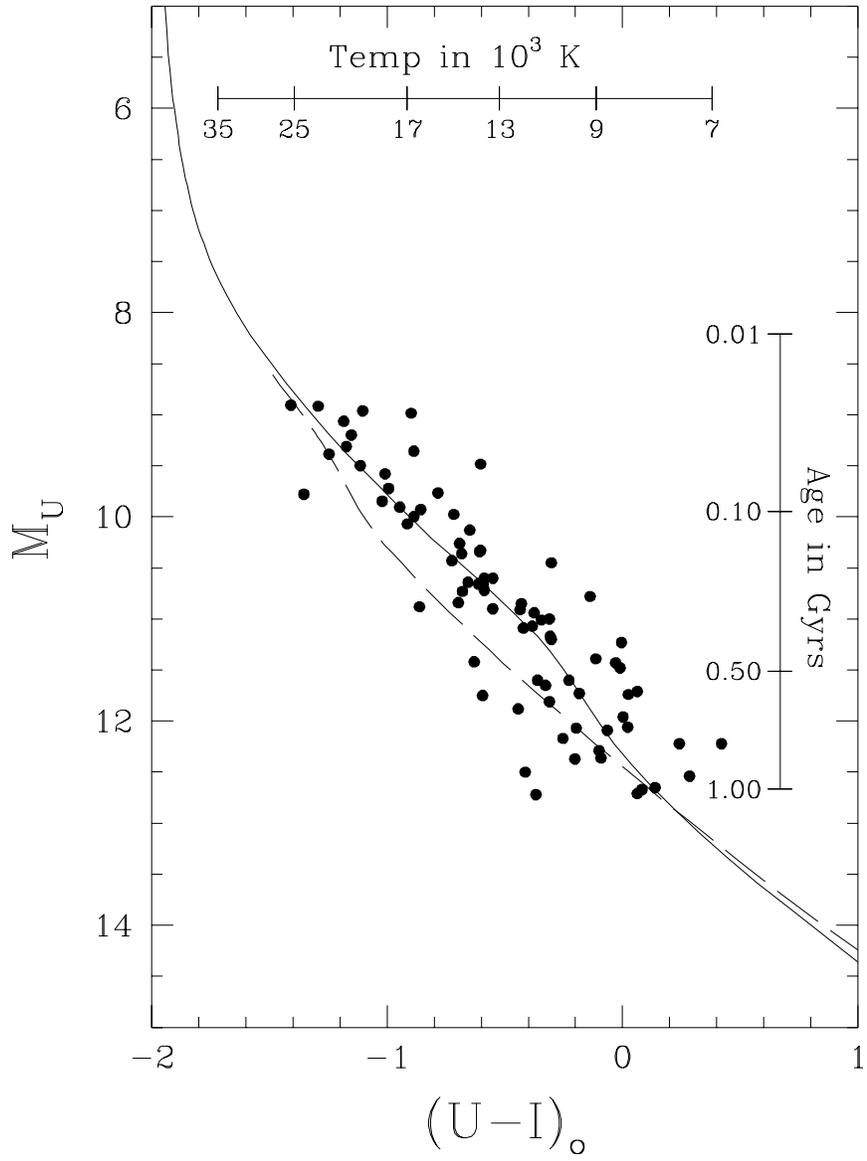

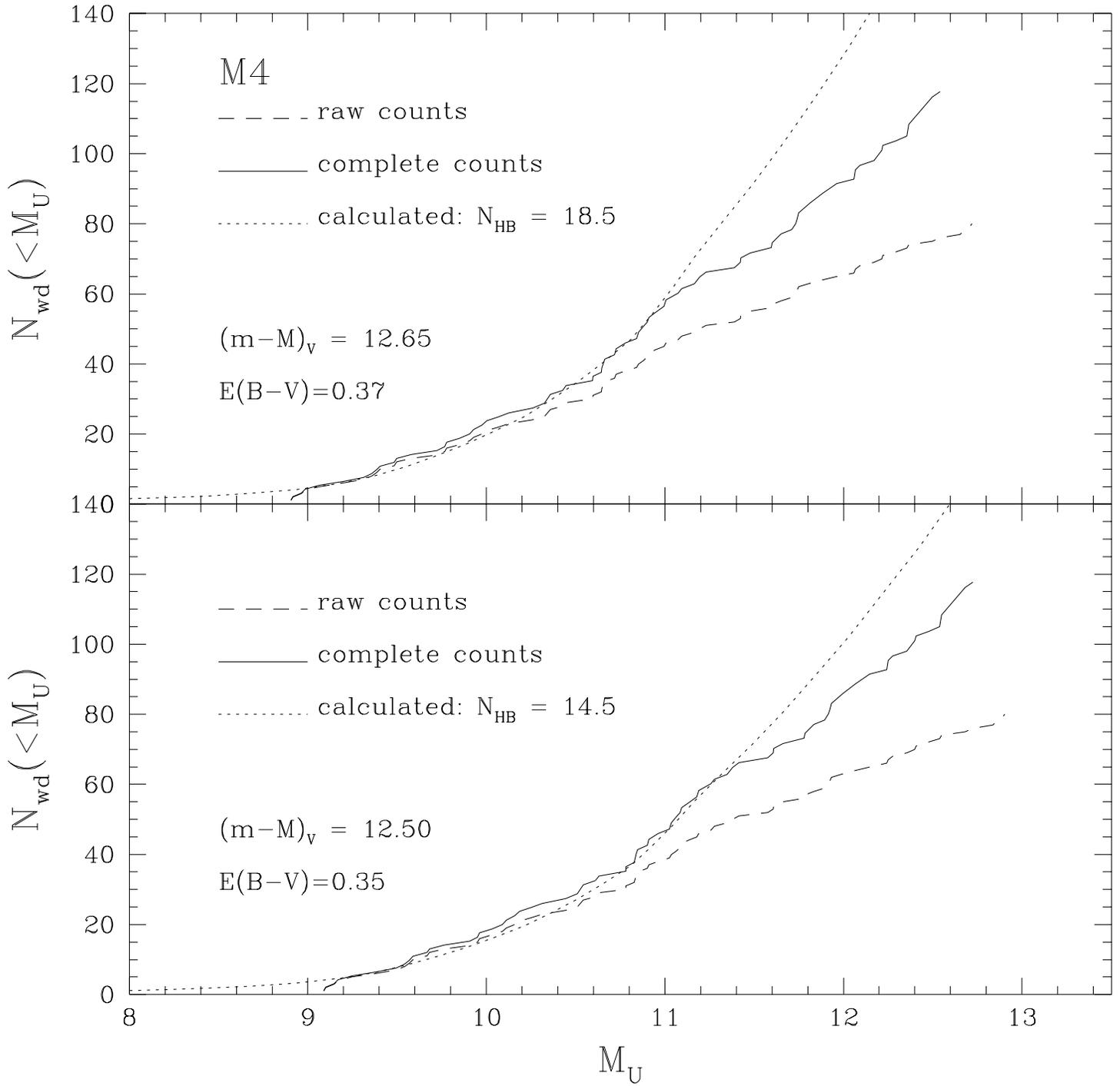

# Hubble Space Telescope Observations of White Dwarfs
# in the Globular Cluster M4[1]

by

Harvey B. Richer[2], Gregory G. Fahlman[2], Rodrigo A. Ibata[2], Peter B. Stetson[3], Roger A. Bell[4], Michael Bolte[5], Howard E. Bond[6], William E. Harris[7], James E. Hesser[3], Georgi Mandushev[2], Carlton Pryor[8] and Don A. VandenBerg[9]

1. Based on observations with the NASA/ESA *Hubble Space Telescope*, obtained at the Space Telescope Science Institute, which is operated by AURA, Inc., under NASA contract NAS5–26555.

2. Department of Geophysics and Astronomy, 129–2214 Main Mall, University of British Columbia, Vancouver, BC, Canada V6T 1Z4. E-mail surname@astro.ubc.ca

3. Dominion Astrophysical Observatory, Herzberg Institute of Astrophysics, National Research Council, 5071 W. Saanich Road, RR5, Victoria, B.C., Canada V8X 4M6. E-mail surname@dao.nrc.ca

4. University of Maryland, Department of Astronomy, College Park, MD 20742–2421. E-mail rabell@astro.umd.edu

5. University of California, Lick Observatory, Santa Cruz, CA 95064. E-mail bolte@lick.ucsc.edu

6. Space Telescope Science Institute, 3700 San Martin Drive, Baltimore, MD 21218. E-mail bond@stsci.edu

7. McMaster University, Department of Physics and Astronomy, Hamilton, ON, Canada L8S 4M1. E-mail harris@physun.physics.mcmaster.ca

8. Department of Physics and Astronomy, PO Box 849, Rutgers, the State University of New Jersey. Piscataway, NJ 08855–0849. E-mail pryor@physics.rutgers.edu

9. University of Victoria, Department of Physics and Astronomy, PO Box 3055, Victoria, BC, Canada V8W 3P7. E-mail davb@uvvm.uvic.ca

**Abstract**

With the Wide Field Planetary Camera 2 (WFPC2) on the Hubble Space Telescope, we have discovered in M4 (NGC 6121, C 1620-264) the first extensive sequence of cooling white dwarfs seen in a globular cluster. Adopting a distance modulus of $(m - M)_V = 12.65$ and a reddening of $E(B - V) = 0.37$, we show that the sequence, which extends over $9 < M_U < 13$, is comprised of white dwarfs of mass $\sim 0.5 M_\odot$. The total mass loss from the present turnoff to the white dwarf sequence is $0.31 M_\odot$ and the intrinsic dispersion in the mean mass appears to be $< 0.05 \ M_\odot$. Both the location of the white dwarf cooling sequence in the cluster color-magnitude diagram and the cumulative luminosity function attest to the basic correctness and completeness of the physics in theoretical models for the upper three magnitudes of the observed white dwarf cooling sequence. To test the theory in globular clusters at cooling ages beyond $\sim 3 \times 10^8$ years will require deeper and more complete data.

subject headings: stars: white dwarfs, globular clusters: M4

## 1. Background, Observations and Reductions

All single stars currently terminating their nuclear burning lifetimes in a globular star cluster are expected to evolve into white dwarf (WD) stars. Indeed, this has been the situation for many billions of years and hence these clusters should be replete with WDs. For example, in a cluster like M4 with a total visual luminosity of $5 \times 10^4 L_\odot$, the expectation is that there should be of order $2 \times 10^4$ WDs present. However, because of their low intrinsic luminosities and despite a number of efforts (Richer 1978, Chan & Richer 1986, Ortolani & Rosino 1986, Richer & Fahlman 1988), no ground-based data have definitively reached the WD population in any globular cluster. Several recent Hubble Space Telescope (HST) results (Paresce et al. 1995, NGC 6397; Elson et al. 1995, $\omega$ Cen; De Marchi & Paresce 1995, M15) have reported small samples of WDs in three clusters, but for these the numbers detected are too small and the photometry too poor to allow the WDs to be used in the sensitive tests of WD structure and evolution discussed below. Our HST observations in the nearby globular cluster M4 were specifically designed to be efficient in locating these very faint, blue objects. An extensive sequence of cooling WDs has been discovered in this system, the closest globular clusters to the Sun.

The observations we discuss below are imaging data taken with the Wide Field Planetary Camera (WFPC2) on the refurbished HST. The PC field was centered at 85.5 arcsec from the cluster center which is slightly larger than one core radius ($r_c$). The orientation of the chips was such that the PC and WF4 were at about the same distances from the cluster center while WF2 and WF3 were both about 150 arcsec from the center. The data that we discuss here consist of eight exposures taken in each of F336W (U, total exposure 11,800 seconds) and F814W (I, total exposure 5500 seconds). Observations have also been obtained in two other fields, one in the cluster core and the other at 4 $r_c$, but we delay the discussion of these data to a more detailed contribution.

The raw data frames had the standard HST pipeline processing performed on them. This includes bias subtraction, correction for dark current and flat-fielding. In addition, hot pixels were flagged and not used in the reduction procedures, vignetted pixels were blanked-out, the charge transfer efficiency correction was made, the decrease in F336W transmission as a function of time from decontamination was accounted for and we corrected for non-uniform illumination of the chips. The photometry was carried out using ALLFRAME (Stetson 1994) with a variable point spread function (PSF) and, for ease of comparison with theoretical results and with WDs whose magnitudes and colors have been measured from the ground, the photometry was transformed to ground-based U and I as defined by the Landolt standards (1983, 1992, 1992a) using the transformations discussed in Holtzman et al. 1995.

## 2. The Cluster Color-Magnitude Diagram

The four panels in Figure 1 display the $M_U, (U - I)_0$ color-magnitude diagrams (CMDs) for the PC field; WF2 and WF3 together since they are at similar distances from the cluster center; WF4, which is the most crowded field and hence has largest photometric errors; and a summed CMD. Using $(m - M)_V = 12.65$ (this derived using an HB luminosity of $M_V(RR) = M_V(HB) = 0.15[Fe/H] + 1.0$ (Carney et al. 1992, Skillen et al. 1993) and $V_{HB} = 13.45$ (Cudworth & Rees 1990)), $E(B - V) = 0.37$ (Richer & Fahlman 1984), $R = 3.1$, $A_U = 4.83E(B - V)$ and $E(U - I) = 2.98E(B - V)$ (Mathis 1990, Bessell & Brett 1988), we transformed the CMD into $M_U$ versus $(U - I)_0$. To be included in these diagrams the stars had to pass two tests. First, they had to be measured by ALLFRAME on all eight U frames and all eight I frames. Secondly, every object was examined by eye to insure that it was in fact a star. A number of spurious objects found their way into the star lists; for example, the crossings of diffraction spikes were often found as stars by the star-finding routines and could only be eliminated by visual inspection. All objects passing both of these criteria were kept in the lists even if their photometry was possibly compromised by being located near a bright star or on a diffraction spike. Main sequence stars brighter than $M_U = 8$ are saturated on our frames in I and brighter than $M_U = 6$ they exhibit bleeding and extended diffraction spikes. Nevertheless ALLFRAME is capable of measuring the magnitudes of such stars by fitting the PSF to their wings. The photometry is, however, less reliable than that for unsaturated objects. This is the reason for the odd appearance of the scatter along the main sequence in Figure 1, where the errors grow both to brighter and fainter magnitudes than $M_U = 8$.

Because of the long color baseline, the WD cooling sequence is prominent and well separated from the main sequence and the background Galactic bulge stars (seen as the scatter of stars around the main sequence) in each of our frames. The onset of the WD cooling sequence is near $M_U = 9$ and it continues to the limit of the data. The faintest WDs detected in F336W are still several magnitudes above the detection threshold on the F814W frames. From the scatter in the measured values on the individual frames, the errors in the magnitudes and colors of the WDs are typically ±0.05 at the bright end of the sequence and ±0.3 at the faint end.

## 3. The Cluster White Dwarfs

### 3.1 The White Dwarf Cooling Sequence: Comparison with Theory

Since recently formed WDs in a globular cluster have all evolved from stars of about the same mass, the location of these WDs in a color-magnitude diagram will fall along a line of essentially constant mass termed the cooling sequence. We have constructed a theoretical cooling sequence for 0.5 $M_\odot$ DA WDs using the pure carbon core evolutionary models of Wood (1995) with helium layers of 1% the mass of the star and thick hydrogen layers of 0.01% of the mass and the DA model atmosphere colors of Bergeron et al. (1996). Also, we calculated a similar sequence for DB WDs with no hydrogen layers and 0.01% helium layers using DB model atmosphere colors (Bergeron et al. 1996). These sequences are shown in Figure 2 together with the M4 WDs. It is clear that the location and shape of the 0.5 $M_\odot$ DA cooling sequence is a good match to most of the data.

We have carried out fits of the M4 WDs to the Wood (1995) cooling curves for DA WDs in the mass range 0.40 to 0.60 $M_\odot$ and find that a mean mass of 0.50 $M_\odot$ provides the best match to the data. A formal error in this mass determination is difficult to assign, but when errors in the cluster properties, the stellar photometry and the transformation from HST photometry to the ground-based system are all included, an uncertainty of about ±0.05 $M_\odot$ is indicated. We caution the reader, however, that the choice of which mass best fits the data depends on the choice of the distance modulus and reddening to the cluster. For example, an uncertainty of 0.1 mag in the distance modulus corresponds to an uncertainty in the mass of the best fitting cooling sequence of about 0.03 $M_\odot$.

### 3.2 The White Dwarf Mass Dispersion

Fusi Pecci & Renzini (1979) suggested that the WD cooling sequence should effectively be dispersionless in a globular cluster, as the horizontal branch (HB) morphology in some globular clusters indicates a very small mass range. Lee (1990) infers that the mass range on the M4 HB is only of the order of 0.03 $M_\odot$. However, Dorman et al. (1989) derive a much larger dispersion for the masses of the 47 Tuc HB stars. To investigate this we constructed artificial DA WD sequences in the magnitude range $M_U = 9.0$ to 12.5 about the best fitting cooling curve. The number of stars used in each sequence was equal to the observed number of candidate WDs. In order to simulate an intrinsic mass dispersion in the WD population, we convolved the artificial sequences in the absolute magnitude direction with Gaussians of standard deviation 0, 0.1, 0.15 and 0.2 mags. As well, the photometric errors as a function of magnitude and color were included. We constructed 1000 such samples for each dispersion and then compared the resultant $\chi^2$ statistic of the fit of the artificial DA WDs to the 0.50 $M_\odot$ cooling curve with that derived for the real stars. We found that $\chi^2$ for the real stars was smaller than 65.6% of the realizations for the zero dispersion sample, smaller than 94.1% of the realizations for the 0.1 mag dispersion sample, less than 99.2% of those for the 0.15 mag input dispersion and smaller than 99.9% of the realizations for the 0.2 mag dispersion. Thus, to within $3\sigma$ accuracy, the observed intrinsic dispersion in magnitude in the M4 WD sample does not exceed 0.15 mags. This corresponds to about 0.05 $M_\odot$, so that the mass dispersion is indeed very small and may be zero. It should be noted that the above simulations will always overestimate the real dispersion as some fraction of the WDs observed in M4 are likely to be non-DAs. For example in the field, 20% of all the WDs with effective temperatures between 30,000 K and 10,000 K are DBs, while DQ and DZ WDs appear in appreciable numbers below 10,000 K. White dwarfs with these different atmospheric compositions are more dispersed in the color-magnitude diagram than a pure DA sample, so much of what we see as scatter may be attributable to composition differences.

Finally we note that the analysis above allows for a new determination of the amount of mass lost in the evolution of a low mass Population II star from the turnoff until it becomes a WD. From the isochrones of Bergbusch & VandenBerg (1992), the turnoff mass in a 14 Gyr globular cluster with [Fe/H] = $-1.26$ is 0.81 $M_\odot$. With M4 WD masses of 0.50 $M_\odot$, the total mass lost per star in the red giant branch, the HB, the asymptotic giant branch and the planetary nebula phases amounts to about 0.31 $M_\odot$.

### 3.3 The White Dwarf Luminosity Function: Comparison with Theory

A prediction for the number of WDs in a globular cluster follows from the conservation of stars as they move rapidly through their post main sequence evolutionary phases. The HB provides a convenient normalization point (Renzini & Buzzoni 1986, Renzini 1988, Fahlman & Richer 1988). Hence, we have

$$N_{WD}(< M_U) = N_{HB}\ T_c(< M_U)/T_{HB} \tag{1}$$

where $N_{WD}(< M_U)$ is the number of WDs brighter than absolute magnitude $M_U$, $T_c(< M_U)$ the WD cooling time to reach $M_U$ and $T_{HB}$ the lifetime on the HB, which we take to be $10^8$ years (Renzini 1977, Dorman 1992). For our data we estimate $N_{HB}$ by counting the number of HB stars in an annulus, whose width is defined by our HST data set, around the center of M4 on available ground-based CCD data. This gives $N_{HB} = 11.7 \pm 3.4$ for the area covered by our HST data where the Poisson uncertainty is indicated. For the comparison between observation and theory, we adopt the DA cooling curves discussed in the previous sections.

In the upper panel of Figure 3 we compare the observed cumulative luminosity function (CLF), both in raw form and with incompleteness corrections applied, to that predicted by equation 1 using the distance modulus and reddening specified in §2. The incompleteness corrections are the inverse of the recovery fractions of artificial stars added into the data frames and are field and magnitude dependent. Individual stars were weighted by their corresponding incompleteness factor to construct the corrected CLF shown as the solid line in Figure 3. The dotted line is the theoretical CLF for the 0.5 $M_{\odot}$ cooling curve with a normalization of $N_{HB}$ = 18.5 in equation 1. This provides a good fit to the data at the bright end ($M_U$ < 11.0 or $T_c < 3 \times 10^8$ years), but lies increasingly above the observations at fainter magnitudes.

The good agreement at the bright end is strong confirmation that cooling theory at the hot end of the WD sequence is well understood. The deviation of theory from observation at the fainter end may reflect some problem with the incompleteness corrections (the recovery fractions are about 0.5 in field 4 and close to 0.75 in the other fields near the departure point at $M_U = 11$ and fall to 0.3 and 0.5 by $M_U$ = 12). In view of the necessity to apply these corrections, and the uncertainty in the composition of the core of the M4 WDs (for example, pure oxygen cores will cool more quickly) and the likely mixture of non-DAs in our sample, it would be premature to draw conclusions from the observed disagreement at the faint end.

The normalization factor ($N_{HB} = 18.5$) is about $2\sigma$ away from the currently observed number of HB stars. This is acceptable, but a little uncomfortable. Better agreement can be obtained by adjusting the distance modulus of M4. For example, adopting $(m-M)_V = 12.50$ (cf. Richer & Fahlman 1984) and $E(B-V)$ = 0.35 will shift the observed CLF by $\delta M_U \sim 0.2$ mags; this leads to a good fit of the observed CLF for stars brighter than $M_U = 11.7$ for $N_{HB} = 14.5$, which is now less than $1\sigma$ away from the current number of HB stars. As more data accumulate, the CLF will itself be a sensitive distance indicator.

## 4. Summary and Conclusions

The observation of an extensive sequence of cooling WDs in the globular cluster M4, covering about 4 mags in $M_U$, has allowed us to make a direct comparison to the predictions arising from the current theory of white dwarf cooling. The similarity between the location and shape of the theoretical cooling curve with that of the WDs in the CMD and the agreement between the calculated luminosity function at the bright end and the observed counts bears strong witness to the basic correctness and completeness of the input physics into the theoretical models. To test the theory at cooling ages beyond about $3 \times 10^8$ years with globular cluster WDs, and in particular to seek the termination of the cluster WD sequence in order to provide a direct measurement of cluster ages, will require deeper and more complete data.

Acknowledgements

Four astronomers provided us with unpublished data and theoretical calculations which made possible most of the analysis undertaken in this contribution. These were Pierre Bergeron, Conard Dahn, Jon Holtzman and Matt Wood. We are indebted to them as much of what has been accomplished in this contribution would not have been possible without their generosity. Karen Schaefer provided excellent support handling the data tapes at the Space Telescope Science Institute. The research of HBR, GGF, WEH and DVB is supported by grants from the Natural Sciences and Engineering Research Council of Canada. Support for work by RAB, MB, HEB, and CP was provided by NASA through grants from the Space Telescope Science Institute, which is operated by the Associated Universities for Research in Astronomy, Inc., under NASA contract NAS5-26555.

**Figure Captions**

Fig. 1: The four panels contain the $M_U, (U-I)_0$ CMD for the PC field, WF2 and WF3 together, WF4 and the summed CMD for fields in the Galactic globular cluster M4. Apparent $U$ magnitudes are indicated along the right-hand ordinate. The PC frame is centered near 1 $r_c$ from the cluster center. The WD cooling sequence is seen as the bluest stars in the diagram stretching from $M_U \sim +9$ to the limit of the data near $M_U \sim +13$.

Fig. 2: The $M_U, (U-I)_0$ CMD for the 1 $r_c$ field in M4 exhibiting the observed cluster WDs together with theoretical cooling sequences for 0.5 $M_\odot$ DA (solid line) and similar mass DB (dashed line) WDs. Along the upper abscissa the effective temperatures of the DA white dwarfs are indicated while along the right-hand ordinate the cooling ages of the DA white dwarfs are shown.

Fig. 3: The cumulative number counts of WDs in the 1 $r_c$ field of M4 plotted against absolute $U$ magnitude. The dotted line represents the counts expected according to equation (1), the dashed line is the raw observed counts while the solid line depicts the incompleteness corrected observed number counts. The two panels are for slightly different cluster distance moduli and reddening.